\documentclass[journal]{IEEEtran}

\usepackage[pdftex]{graphicx}
\usepackage{amsmath}
\usepackage{url}

\begin{document}
\bstctlcite{IEEEexample:BSTcontrol}
\pagestyle{empty}
\title{Online Calibration of the TPC Drift Time in the ALICE High Level Trigger}
\pagestyle{empty}
\author{David Rohr,
        Mikolaj Krzewicki,
        Chiara Zampolli,
        Jens Wiechula,
        Sergey Gorbunov,
        Alex Chauvin,
        Ivan Vorobyev,
        Steffen Weber,
        Kai Schweda,
        Volker Lindenstruth
        for the ALICE Collaboration
\thanks{Manuscript received June 25, 2016; revised January 23, 2017}
\thanks{David Rohr, Frankfurt Institute for Advanced Studies, CERN, e-mail: drohr@jwdt.org.
Mikolaj Krzewicki, Sergey Gorbunov, and Jens Wiechula, Goethe University Frankfurt.
Chiara Zampolli, INFN, CERN.
Kai Schweda, University of Heidelberg.
Ivan Vorobyev and Alex Chauvin, Technical University of Munich.
Steffen Weber, Technical University Darmstadt, GSI.
Volker Lindenstruth, Goethe University Frankfurt, Frankfurt Institute for Advanced Studies, GSI.}%
}
{}
\pagestyle{empty}
\maketitle
\thispagestyle{empty}
\pagestyle{empty}

\begin{abstract}
ALICE (A Large Ion Collider Experiment) is one of four major experiments at the Large Hadron Collider (LHC) at CERN.
The High Level Trigger (HLT) is a compute cluster, which reconstructs collisions as recorded by the ALICE detector in real-time.
It employs a custom online data-transport framework to distribute data and workload among the compute nodes.

ALICE employs subdetectors sensitive to environmental conditions such as pressure and temperature, e.g. the Time Projection Chamber (TPC).
A precise reconstruction of particle trajectories requires the calibration of these detectors.
Performing the calibration in real time in the HLT improves the online reconstructions and renders certain offline calibration steps obsolete speeding up offline physics analysis.
For LHC Run 3, starting in 2020 when data reduction will rely on reconstructed data, online calibration becomes a necessity.
Reconstructed particle trajectories build the basis for the calibration making a fast online-tracking mandatory.
The main detectors used for this purpose are the TPC and ITS (Inner Tracking System).
Reconstructing the trajectories in the TPC is the most compute-intense step.

We present several improvements to the ALICE High Level Trigger developed to facilitate online calibration.
The main new development for online calibration is a wrapper that can run ALICE offline analysis and calibration tasks inside the HLT.
On top of that, we have added asynchronous processing capabilities to support long-running calibration tasks in the HLT framework, which runs event-synchronously otherwise.
In order to improve the resiliency, an isolated process performs the asynchronous operations such that even a fatal error does not disturb data taking.
We have complemented the original loop-free HLT chain with ZeroMQ data-transfer components.
The ZeroMQ components facilitate a feedback loop, that after a short delay inserts the calibration result created at the end of the chain back into tracking components at the beginning of the chain.
All these new features are implemented in a general way, such that they have use-cases aside from online calibration.

In order to gather sufficient statistics for the calibration, the asynchronous calibration component must process enough events per time interval.
Since the calibration is only valid for a certain time period, the delay until the feedback loop provides updated calibration data must not be too long.
A first full-scale test of the online calibration functionality was performed during the 2015 heavy-ion run under real conditions.
Since then, online calibration is enabled and benchmarked in the 2016 proton-proton data taking.
We present a timing analysis of this first online-calibration test, which concludes that the HLT is capable of online TPC drift time calibration fast enough to calibrate the tracking via the feedback loop.
We compare the calibration results to the offline calibration and present a comparison of the residuals of the TPC cluster coordinates with respect to offline reconstruction.
\end{abstract}


\IEEEpeerreviewmaketitle

\section{Introduction}

\label{sec:scheme}

\IEEEPARstart{A}{LICE} is one of the four large-scale experiments at the Large Hadron Collider (LHC) at CERN~\cite{bib:alice}.
Its main purpose is the study of matter under extreme conditions of high temperature and pressure.
This is done through the collisions of lead nuclei accelerated by the LHC to the highest energies possible today.
The design lead-lead interaction rate inside ALICE is 8\,kHz.
The ALICE detector also takes data from proton-proton collisions, both as a reference for lead-lead collisions and for other studies on their own.

\looseness=-1
The ALICE High Level Trigger (HLT)~\cite{bib:alice-hlt} is an online compute farm for real time processing of the events recorded by ALICE.
Based on the reconstruction of the data in real time, the HLT can trigger for or tag interesting events.
On top of that, the HLT performs a compression of the raw data.
During normal operation, the HLT receives and processes an incoming data rate of up to 30\,GB/s.
The most compute intense part of event reconstruction is the reconstruction of particle trajectories, called tracking.
The most important detectors with respect to tracking are the Time Projection Chamber (TPC) and the Inner Tracking System (ITS).
For the TPC track reconstruction, the HLT employs a GPU-accelerated algorithm based on the Cellular Automaton and the Kalman Filter~\cite{bib:tns,bib:chep}.

Several detectors of the ALICE experiment are sensitive to environmental conditions such as temperature and pressure.
The environmental conditions influence several characteristics of the detectors like gain or drift time.
In order to perform a precise reconstruction, the detector calibration must take these effects into account.
Since environmental conditions change during data-taking, an initial calibration is insufficient.
Instead, the detectors must be calibrated continuously.

\begin{figure}[!b]
\centering
\includegraphics[width=3.0in]{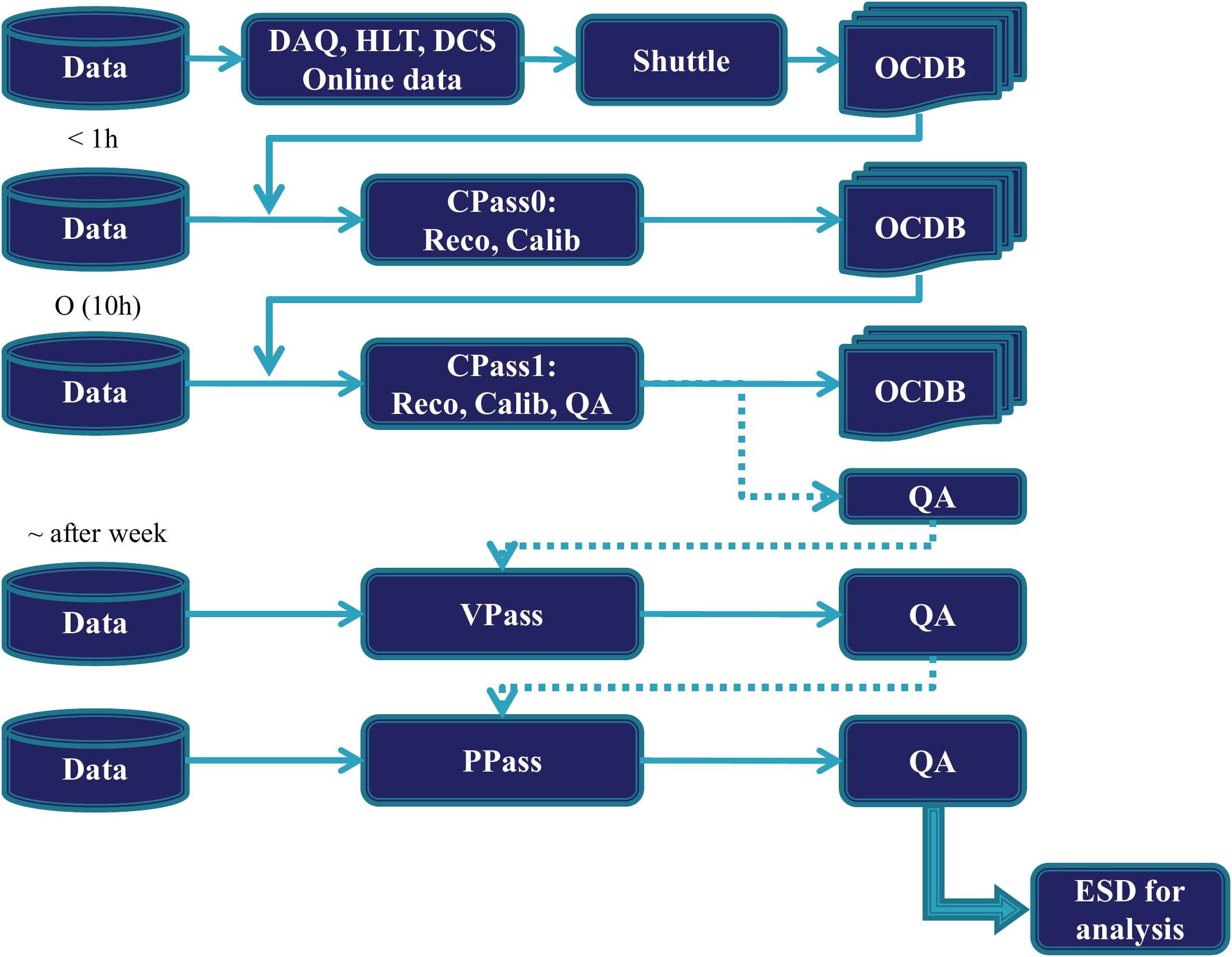}
\caption{Illustration of the reconstruction passes used in ALICE offline analysis to produce calibration objects. (From Ivan Vorobyev: Online Calibration of the ALICE-TPC in LHC-Run 2, DPG Fr\"uhjahrstagung 2015.)}
\label{fig_alice-calib}
\end{figure}

\looseness=-1
Sensors alone, e.\,g.~for pressure and temperature, are not enough for detector calibration.
Instead, the calibration needs the tracks of reconstructed events.
For instance, TPC drift time is calculated from the misalignment of matched reconstructed TPC and ITS tracks.
This imposes a cyclic dependency: the reconstruction needs the calibration which in turn relies on the reconstruction.
Normal ALICE Offline reconstruction copes with this via multiple reconstruction and calibration passes for each run.
The first offline pass is the CPass 0, which reconstructs a subset of the events that match a calibration trigger using default calibration objects.
Then, the CPass 0 calibration objects including the TPC drift time object are created for that run and stored in the OCDB (Offline Conditions DataBase).
The next pass is the CPass 1, which reconstructs the data again using the TPC drift time calibration created in CPass 0.
Afterwards, the other calibration tasks produce the remaining calibration objects.
This two-stage procedure is necessary because the calibration of other detectors requires a calibrated TPC first.
The third pass consists of a verification whether the calibration is correct and the final reconstruction for physics analysis.
Figure~\ref{fig_alice-calib} illustrates the offline procedure.
This approach is infeasible in the HLT: it is impossible to keep all the data in memory to process the entire data set multiple times.

As a first step to investigate solutions for online calibration in the HLT, we attempt to perform the CPass 0 calibration for the TPC drift time online.
The approach is general enough to be applied to additional calibrations later on, as long as they match certain time constraints and there is sufficient compute capacity in the HLT.

\begin{figure*}[!t]
\centering
\includegraphics[width=7.1in]{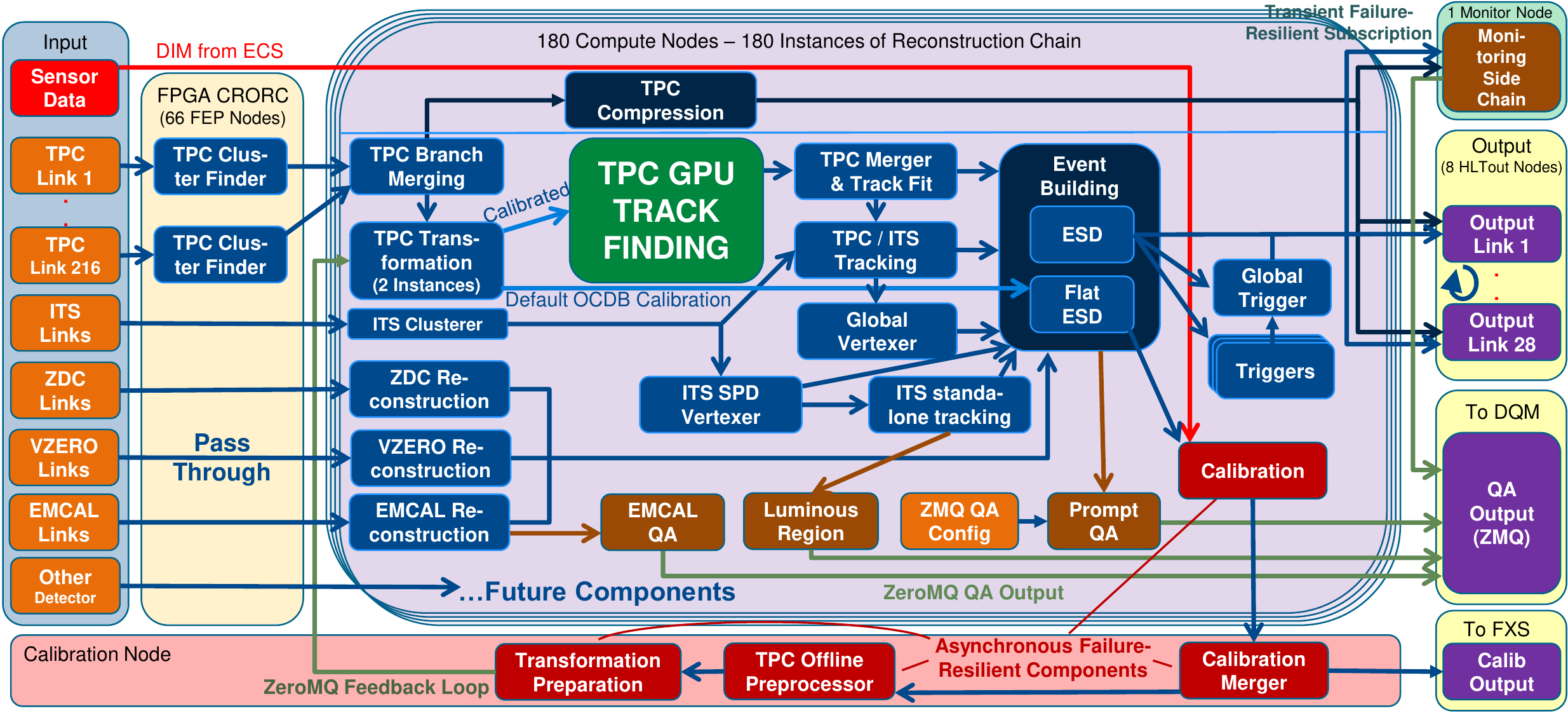}
\caption{Overview of all components currently running in the HLT. \cite{bib:ctd}}
\label{fig_hlt_overview}
\end{figure*}

Processing in the HLT is based on components running in a custom framework based upon the publisher subscriber principle.
There are source components that feed the data from the detectors into the processing chain.
Sink components at the end of the chain send the HLT output to data acquisition (DAQ).
In between there are processing components, which receive input from all the components they subscribe to, and provide output to subscribers.
One example is the TPC track finder which subscribes to the TPC clusters as input and provides the TPC tracks as output.
The HLT can be seen as a directed graph where the components are nodes and the subscriptions are edges.
All the components process the incoming events in a pipeline, the framework automatically gathers all required inputs for one event and provides them to a processing component.
By definition, this prohibits circular dependencies, thus the HLT graph is without loops.
Figure~\ref{fig_hlt_overview} in the next section gives an overview of all components currently running in the HLT.

\section{Online Calibration}

\begin{figure}[!b]
\centering
\includegraphics[width=2.5in]{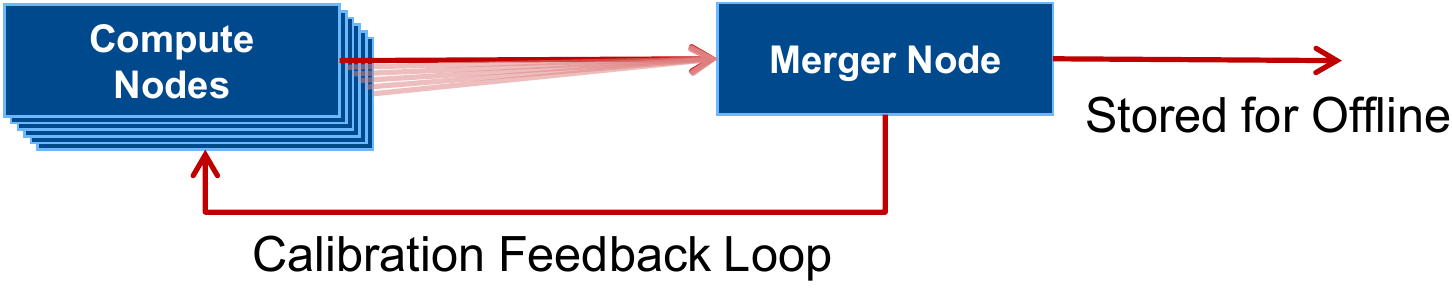}
\caption{Illustration of the scheme for merging calibration objects and for the feedback loop. \cite{bib:ctd}}
\label{fig_calib_merger}
\end{figure}

\begin{figure*}[!t]
\centering
\includegraphics[width=6.0in]{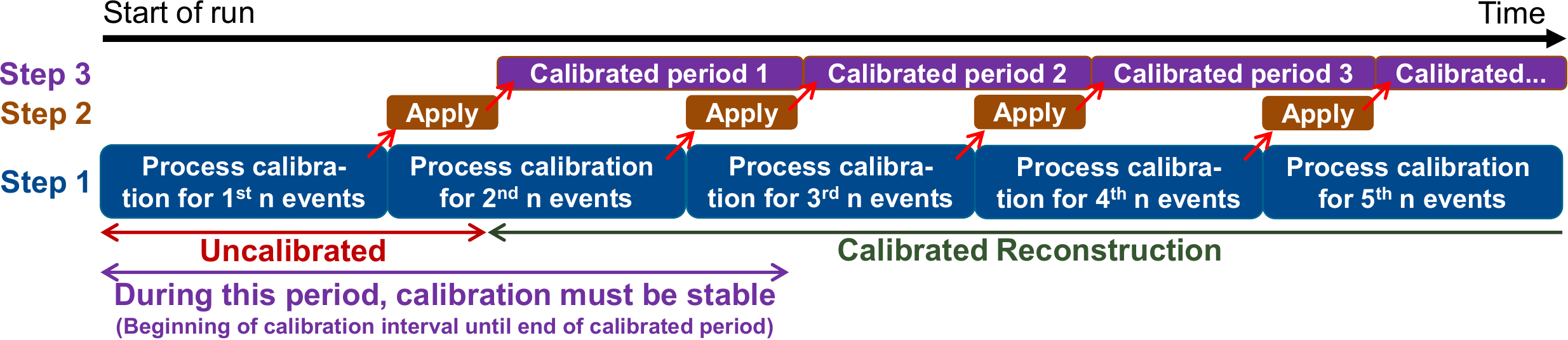}
\caption{Illustration of the online calibration approach in the ALICE High Level Trigger. \cite{bib:rt2016}}
\label{fig_calib}
\end{figure*}

Online calibration in the High Level Trigger has several advantages:
\begin{itemize}
\item The online calibration results improve the online reconstruction in the HLT.
\item There are more elaborate possibilities for online QA when the calibration is already available during data-taking.
\item When the online calibration results are sufficiently good, it is no longer needed to perform certain calibration tasks offline, which can reduce the offline compute requirements.
\item Future experiments like ALICE during LHC Run 3 or the FAIR experiments at GSI will face significantly higher data rates.
    It will be infeasible to store the entire raw data for later analysis.
    Instead, more elaborate triggering and data compression algorithms are mandatory, which need precise online reconstruction.
\end{itemize}

Whereas calibration results are not stable through an entire data taking run, they are stable for a certain time period.
In case of the TPC drift time calibration, which we want to discuss exemplarily, the stability interval is defined by the fastest possible weather change, and we assume that a calibration computed at a certain point in time is valid for at least the following 15 minutes.
This opens a possibility for real time calibration in the HLT.
As the first offline reconstruction pass, the HLT starts with a default calibration.
It processes as many events as needed to gather sufficient statistics for the calibration.
This statistic is gathered by many instances of the calibration tasks spread among all nodes of the HLT cluster to speed up the process.
As a second step, when enough statistics are aggregated, the calibration results from all tasks are collected and merged.
The feedback loop (see Section~\ref{sec_loop}) updates the calibration objects used for the reconstruction in the cluster.
In the third step, the calibration is applied to the reconstruction for all newly processed events.
The final, last calibration objects created contain the entire statistics of all calibration intervals of the run.
This object is stored for offline usage.
Figure~\ref{fig_calib_merger} illustrates the merging procedure and the feedback loop and Figure~\ref{fig_calib} gives a general overview of the calibration process.
One compute node is dedicated to the merging of calibration objects and to preprocessing stages of the feedback loop.
The three steps are arranged in a pipeline.
As soon as the tasks have shipped their current calibration objects to the calibration merger, they restart with the computation of the next calibration immediately.
By and large, the calibration created for the events recorded during the first couple of minutes of the run is used for the events recorded during the following minutes, as long as the calibration is still valid.
While the calibration is applied for several minutes, the pipeline already prepares the calibration for the next couple of minutes.
There is an uncalibrated period at the beginning of a run before the first calibration objects are distributed in the cluster and applied.
Afterwards, the reconstruction of incoming data uses the calibration.
This approach is feasible if the total time from the start of a calibration period (first blue, lower box) until the end of the calibrated period (first purple, upper box) is shorter than the stability interval of the calibration, 15 minutes for the TPC drift time.
The next calibration object is used as soon as it is ready.
To estimate the stability interval, the TPC group assessed how fast the parameters can change, that affect the drift time.
The significant driving factor is the pressure.
Assuming a fast total wheather change, $15$ minutes is the interval where the influence of the pressure change is still in the shadow of the intrinsic TPC resolution.

\subsection{Feedback Loop}

\label{sec_loop}

The purpose of the feedback loop is to distribute the new calibration objects in the cluster and update the objects used by the reconstruction.
Since the HLT data transport framework must be loop-free by design, we cannot use it for the feedback loop.
Thus, we have implemented new HLT components that provide asynchronous ZeroMQ based data transport, which we use for the feedback loop.
In contrast to the normal HLT data flow, this new transport is not used for the distribution of event data, which must be processed in an event synchronous way.
The asynchronous transport feeds back the calibration result at some later point in time, when the events used to generate the calibration have left the HLT already.

The HLT TPC reconstruction does not use the drift time calibration object directly.
The most important processing step of the HLT that involves calibration is the transformation of the TPC clusters from TPC internal coordinates (padrow, pad, and time) into spacial coordinates.
This transformation is time consuming taking into account all effects in the TPC, thus the HLT precalculates a fast transformation map that interpolates the transformation with splines.
The map is precalculated by sampling the normal (offline) transformation in the TPC volume.
When the feedback loop updates the calibration, it must recalculate this map.
This calculation is performed only once on the dedicated calibration node to save compute resources.
These objects are then distributed in the cluster.

The usual access to configuration (calibration) objects is either file based or through alien\footnote{\raggedright \url{alien.web.cern.ch}, \url{aliweb.cern.ch/Offline/Activities/ConditionDB.html}}, which is too slow for HLT operation.
The HLT transfers updated configuration objects in memory.
We have improved the configuration manager such that it can update configuration objects with in-memory copies on the fly.
The TPC cluster transformation was adapted to reconfigure itself with updated configuration objects over time.
The update of configuration objects on-the-fly is implemented in a general way such that it can be used for other objects as well later on.

\subsection{Requirements}

Performing the calibration online in the HLT poses several prerequisites.
We have developed several new components for the HLT to solve these requirements.
We did not want to modify the HLT transport framework itself, which has proven its stability during several years of data taking.
Therefore, we implemented all the new features as processing components.
The components are as general as possible, such that they can be applied for other purposes besides online calibration.
In the following, we briefly list these requirements and give references to the previous efforts.
\begin{itemize}
\item The calibration requires a fast track reconstruction algorithm for its input.
We use a fast GPU-accelerated component for the track reconstruction inside the TPC, which is the computationally most expensive part of reconstruction~\cite{bib:tns,bib:chep,bib:cnna}.
\item Originally, ITS tracking in the HLT was performed by means of extrapolating TPC tracks inside the ITS.
This approach does not suit the needs of TPC ITS matching in the calibration because it introduces a bias.
The calibration task needs ITS standalone tracks.
Full ITS standalone tracking is very compute intense, in particular in dense lead-lead events, because the density at the center of the collision where ITS is located is much higher, and because ITS has fewer layers which complicates the combinatorics of track seeding.
As a solution, we employ a new fast ITS standalone tracker whose focus lies on speed and tracking resolution but not on efficiency~\cite{bib:ctd}.
It is not necessary to find all ITS tracks for the matching in the calibration.
\item New asynchronous processing components enable the execution of long-running tasks in the HLT without stalling the event synchronous processing chain~\cite{bib:chep2015}.
The processing is performed in an isolated process.
Even fatal errors like segmentation faults affect only this process and can thus not interfere with ALICE data taking.
\item A new ZeroMQ based data transport mechanism facilitates the feedback loop that distributes the calibration result in the cluster~\cite{bib:chep2015}.
This new transport mechanism is, for instance, also used to send histograms to the data quality monitoring system (DQM) for quality assurance.
\item A new flat data structure provides fast access to reconstructed events to all HLT components avoiding ROOT serialization yielding a speedup of a factor of~$10$ over the normal offline data structure~\cite{bib:ctd}.
\item The feedback loop needs a mechanism to distribute new calibration objects and update the objects in use by the reconstruction code (see Section~\ref{sec_loop}).
\item Even though the feedback loop enables the creation of calibrated TPC clusters for the tracking, the calibration component itself still needs uncalibrated clusters for the TPC-ITS matching.
Our solution is described in Section~\ref{sec:dep} in a broader context.
\end{itemize}

\subsection{Implementation}

In order to execute the actual calibration software in the HLT we have created a new component.
This new component is a wrapper for standard ALICE analysis tasks, which can run any task that supports the new flat data structure~\cite{bib:ctd}.
In this way, we can run the same code in the HLT that is used for the calibration offline.
Running as an isolated process of the asynchronous component, a failure in the offline calibration code has no effect on normal HLT operation.
If the process crashes, the current event is not used for the calibration and the process is restarted continuing with the next event.
Figure~\ref{fig_hlt_overview} gives an overview of all components running in the HLT.
The newly developed components for calibration, based on the new asynchronous failure-tolerant component support, are shown in red.
The asynchronous component does not process all events synchronously with the normal HLT chain, but it skips incoming events when its isolated worker process is busy.
It processes as many events as possible on a best effort principle, but under load usually operates at a slower rate than the reconstruction.
An input queue ensures that the worker process is always busy.

For offline operation, ALICE creates several configuration objects after the run that contain graphs with sensor readings, e.\,g.~for pressure and temperature, which are used for the calibration (Figure~\ref{fig_alice-calib}).
Some of this information, e.\,g.~the pressure sensors, are available at runtime via the DCS (detector control system) DIM server.
Others are detector-internal and currently not available at runtime.
The HLT does not yet support updates to these objects at runtime, but this can be added in the same manner as for the drift time calibration update.
Hence, the HLT creates constant objects with the sensor information that is available at the start of the run, but without the time dependency, and provides this to all its components.
Only default objects are available for sensors not accessible at runtime.
We will discuss the implications of this aspect later.

\begin{figure*}[!t]
\centering
\includegraphics[width=6in]{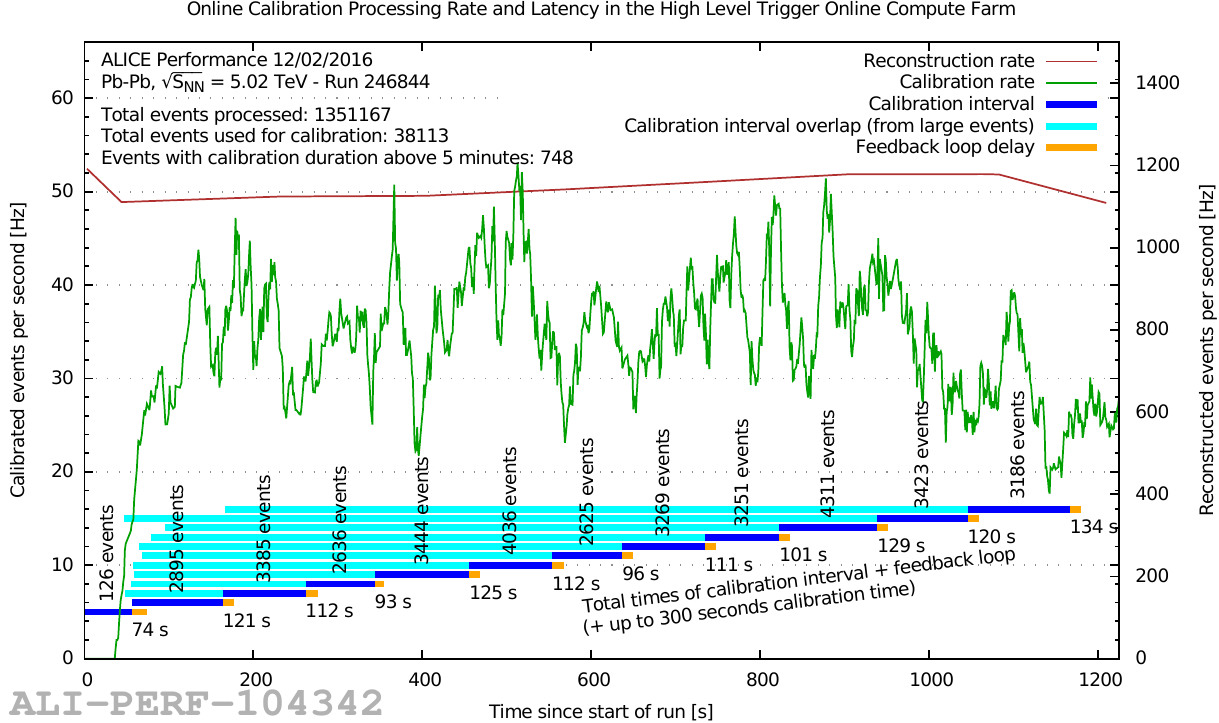}
\caption{Online calibration processing rate and latency in the High Level Trigger online compute farm.
The red (upper) curve shows the rate at which the HLT reconstructs the incoming events synchronously (right-hand scale).
The green (lower) curve shows the asynchronous rate (left-hand scale) how many calibration tasks have finished processing an event in that second.
The bar plot on the bottom show the calibration intervals.
The dark (blue) bar is the calibration interval itself.
The yellow bar on the right of the calibration interval shows the time required for the feedback loop.
The light blue bar on the left of it shows when the first event was recorded that contributed to the calibration interval.}
\label{fig_calib_rate}
\end{figure*}

\section{First Real Test of Online Calibration and Analysis}

We performed a full-scale test of all the new features under real conditions during the December 2015 lead-lead data taking.
At lead-lead event sizes, around 5000 events are sufficient to gather the statistics needed for the TPC drift time calibration.
In this test we employed 120 compute nodes with 5 CPU cores per node for the calibration task.
Figure~\ref{fig_calib_rate} shows the processing rate of the calibration tasks.
While the average HLT event reconstruction rate has been around 1.15\,kHz, the calibration components processed in average 31 events per second.
In the meantime, by using an improved configuration and using all HLT compute nodes, we improved the calibration processing rate of the December 2015 lead-lead data to 81\,Hz.
Specifically, the green (lower) line shows the number of events that finished the calibration in that second.
Due to the asynchronous nature of the component, the calibration task can have been running for a longer time before.
Therefore, the calibration rate is 0\,Hz at the very beginning of the run because all events take at least a few seconds.
The curve is noisy because the calibration time varies greatly and the rate is not high enough for this to average out.
As described in Section~\ref{sec:scheme}, the online calibration with feedback loop works if the total sum of the time of the calibration period, the time required to apply the feedback loop, and the period during which the calibration is active remains below 15 minutes.
The times are visualized as the bars at the bottom of Figure~\ref{fig_calib_rate}.
The dark (blue) bar represents the calibration interval of usually around 2 minutes.
The light (yellow) bar on the right represents the time required for the feedback loop.
The only significant contribution to the feedback loop comes from the preparation of the new TPC transformation map, which takes 11 seconds in average and thus plays only a minor role.
The calibration is then applied until the next feedback loop provides a new transformation map, i.\,e.~after 2 minutes like the calibration interval.
However, due to the asynchronous component, we also have to consider when the asynchronous calibration tasks started that end up in a calibration interval.
The light (blue) on the left bar shows when the first event was taken that contributed to a respective calibration interval.
This additional delay can be up to 30 minutes for the last calibration interval, and is thus too long to fit in the stability interval of 15 minutes for the feedback loop.
Fortunately, only the largest events take that long.
In total, only 2\,\% of the events need more than 5 minutes for the calibration.
Therefore, by using only the 98\,\% of smaller events, the total maximum time for calibration and feedback loop is $5 + 2 + 2 + 0.2 = 9.2$ minutes and thus shorter than the stability interval.
On top of that, the large events that take too long are not ignored, but their data are added to the previous calibration period where they belong to.
Thus, they contribute to the final calibration object stored for offline but not to the feedback loop.
This demonstrates that the HLT is capable of processing the online calibration with feedback loop in the scenario with the highest load.

\begin{figure}[!b]
\centering
\includegraphics[width=3.4in]{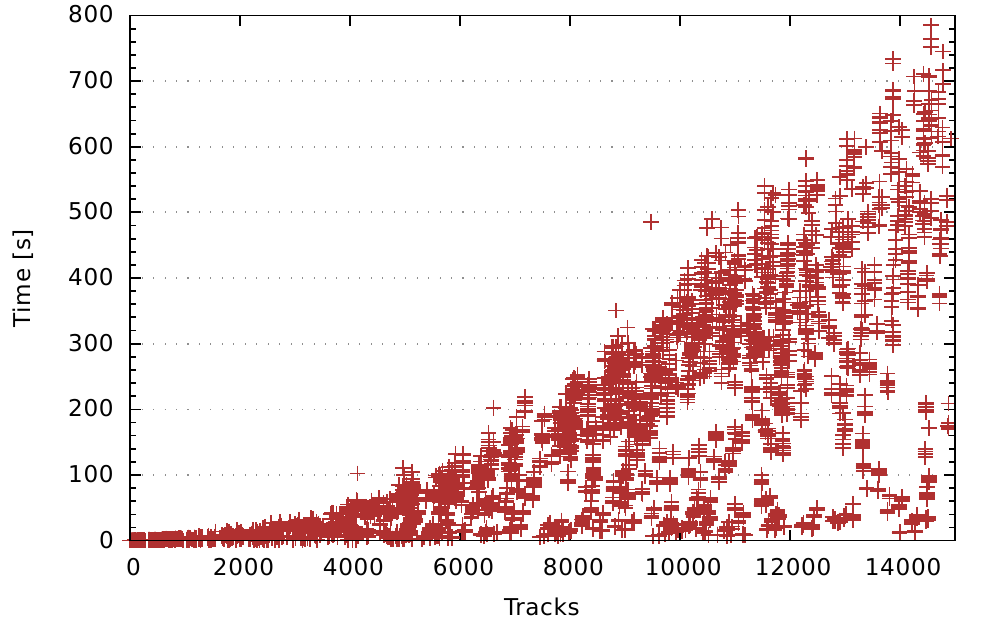}
\caption{Processing time of the calibration task in the HLT depending on the number of tracks in event.}
\label{fig_calib_time}
\end{figure}

Figure~\ref{fig_calib_time} investigates the dependency of the processing time on the event size in terms of number of tracks.
The time required for the calibration grows superlinearly with the number of tracks and over a large range quadratically.
The calibration needs to match a certain number of TPC and ITS tracks.
This can be facilitated faster by processing more smaller events than few larger events.
Therefore, we apply a cut for a maximum of few thousand tracks per event.
The exact limit is to be fine tuned.
By applying a small cut, we can decrease the total time to significantly less than 15 minutes.
On top of that, we have in the meantime improved the matching strategy speeding up the calibration task by a factor of three without affecting the result.
We are also investigating faster propagation algorithms, which offer a speedup of an order of magnitude at marginally reduced accuracy -- similarly to the fast GPU tracking in~\cite{bib:tns}.
In exchange, we have fixed a bug in the track refit used for the 2015 online calibration, which increases the CPU load for the new data.
We need further analysis to find the best trade off in this respect, but in total we can reduce the total compute capacity requirement significantly.
Shortening the time until the availability of calibration results makes the approach feasible for other calibration tasks with shorter stability intervals.

\begin{figure}[!t]
\centering
\includegraphics[width=3.4in]{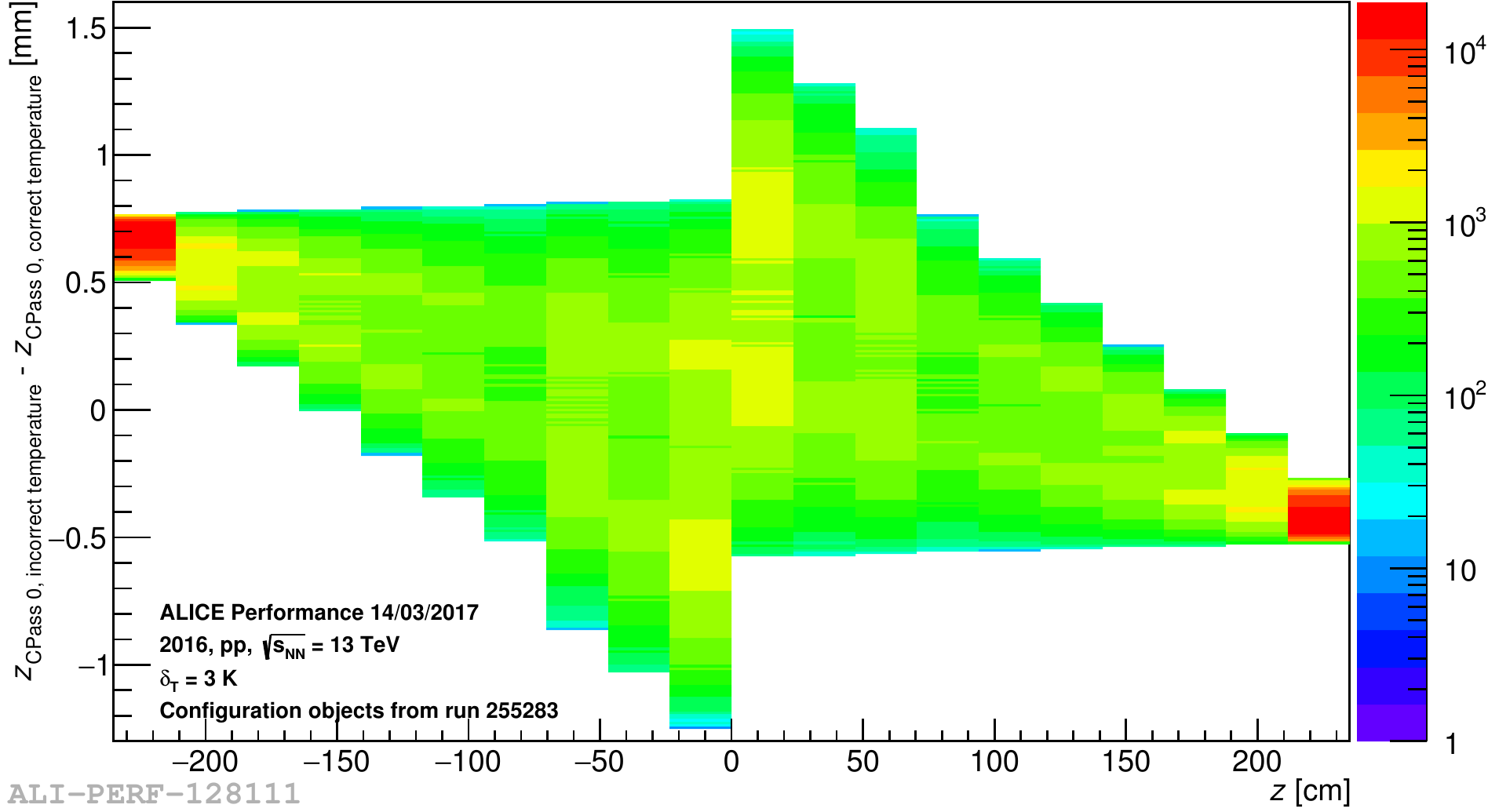}
\caption{Deviation of cluster $z$-position due to missing temperature information in the HLT after drift time correction.
The deviation on the~$y$-axis is obtained as follows: calibration for the run is performed twice, once with a correct temperature setting and once with an offset.
We want to verify that the drift time correction factor from the calibration compensates for the incorrect temperature.
We perform the cluster transformation twice, with both calibrations.
The difference shown on the~$y$-axis is the cluster-per-cluster difference of the clusters'~$z$-coordinate with both calibrations.
}
\label{fig_calib_temperature}
\end{figure}

In the following, we compare the quality of the online calibration objects using the current software to the offline CPass 0 calibration.
We do this by means of comparing the deviation of the clusters' {$z$-position} after the cluster transformation, and by comparing the drift velocity correction factor on its own.
It must be taken into account, that the sensor information available for pressure, temperature, etc.~are different for the online calibration case and the offline CPass 0 case.
Currently, the online calibration knows the pressure at start of run but it does not receive updated pressure information over time.
There is no temperature information yet, thus, the online calibration has to stick to default temperature values.
In the offline CPass 0 case, the time evolution of all these properties is available.
The drift velocity~$v_D$ is computed as $v_D(t) = (1 + \alpha(t)) \cdot v_0(T, p, t)$, where~$v_0(T, p, t)$ is the theoretical drift velocity for a given temperature~$T$ and pressure~$p$ whereas~$\alpha(t)$ is the time dependent correction factor calculated by the drift time calibration~\cite{bib:mikolaj}.
Pressure and temperature affect the default drift velocity~$v_0$, but an incorrectly computed~$v_0$ should be compensated for by the correction factor~$\alpha$ if the effect is homogeneous inside the TPC.
The drift time calibration object that contains~$\alpha$ will be different, depending on which temperature and pressure information was used for~$v_0$ in the calibration task.
However, the effect on the final cluster~{$z$-position} should be small.
It is only important to use the same pressure and temperature objects in the reconstruction that were used during the drift time calibration.
We analyse the effect of incorrect temperature information in Figure~\ref{fig_calib_temperature}.
It shows the deviation of the cluster-{$z$-position} calculated with the correct and with an incorrect temperature on a per-cluster basis -- in both cases applying the correct drift time calibration calculated for the respective temperature.
The drift time calibration computes the best fit of the drift velocity to match TPC and ITS tracks, which results in an average~$z$-deviation close to~$0$.
The deviation for individual clusters is below~$1.5$\,mm.
This is small enough to use the online calibration today but it is also an aspect that we want to address in the future.
We have used a configuration object with an incorrect temperature differing by around~$3$\,K, which is a large difference that will not appear in reality but can be seen as a worst case study.
The online calibrated~$z$-deviations of the HLT clusters shown in the following figures are thus much smaller.

\begin{figure}[!t]
\centering
\includegraphics[width=3.36in]{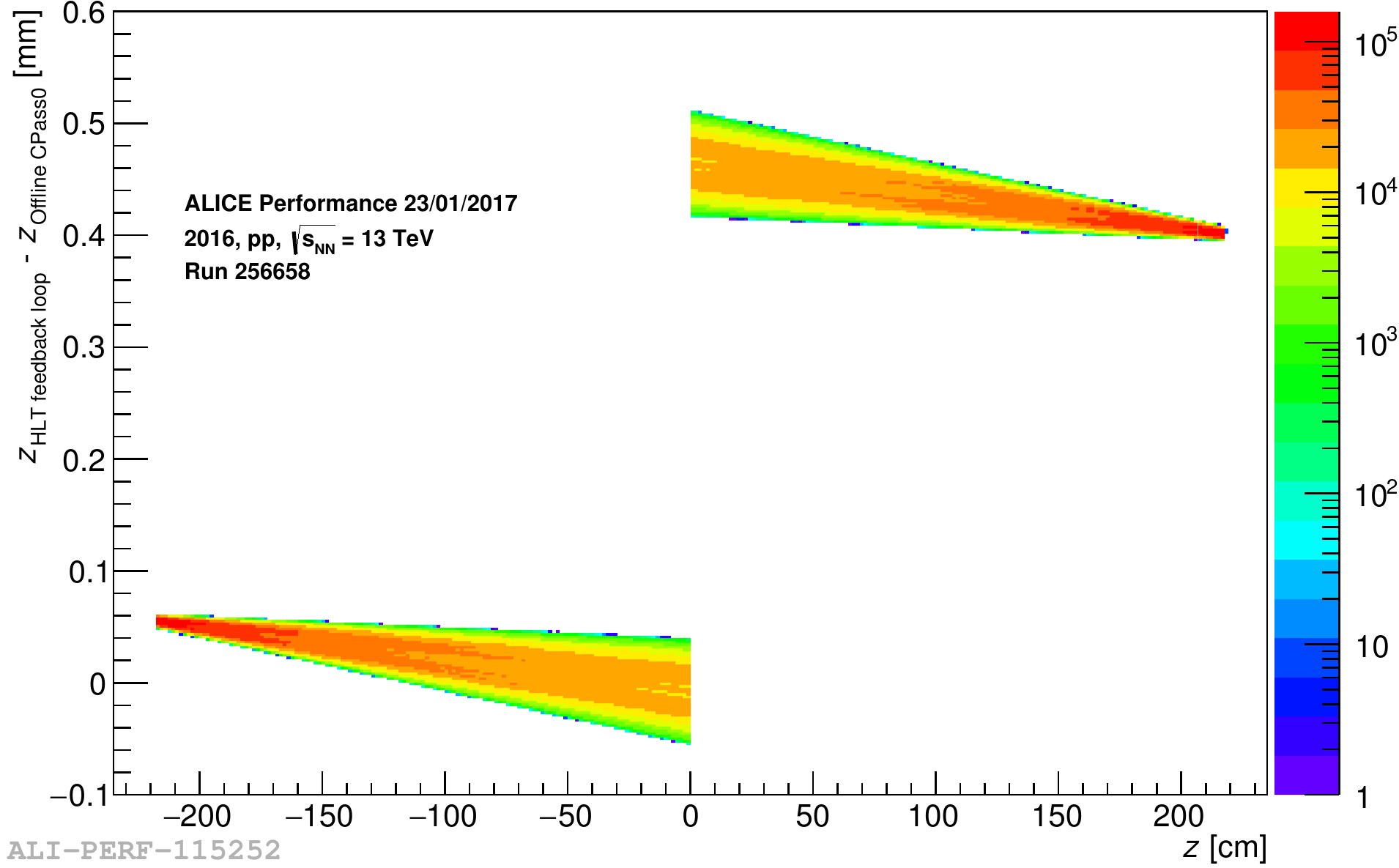}
\caption{Deviation of cluster $z$-position using HLT online calibration with feedback loop compared to offline reconstruction.
Similar to Figure~\ref{fig_calib_temperature}, but on the~$y$-axis we compare the difference of the clusters'~$z$-position obtained using the offline calibration and using the online calibration.
(The difference between positive and negative~$z$ stems from a currently incorrectly calculated trigger offset time in the online calibration.)}
\label{fig_calib_dz}
\end{figure}

Next, we compare the deviation of the transformed cluster coordinates in the HLT using online calibration to the cluster coordinates obtained offline after CPass 0.
Figure~\ref{fig_calib_dz} shows the deviation of the~{$z$-position} versus~$z$, which is proportional to the drift time.
The error after the projection on the~$z$-axis is bigger close to the central electrode at~$z = 0$ due to local distortions in the TPC.

\begin{figure}[!t]
\centering
\includegraphics[width=3.4in]{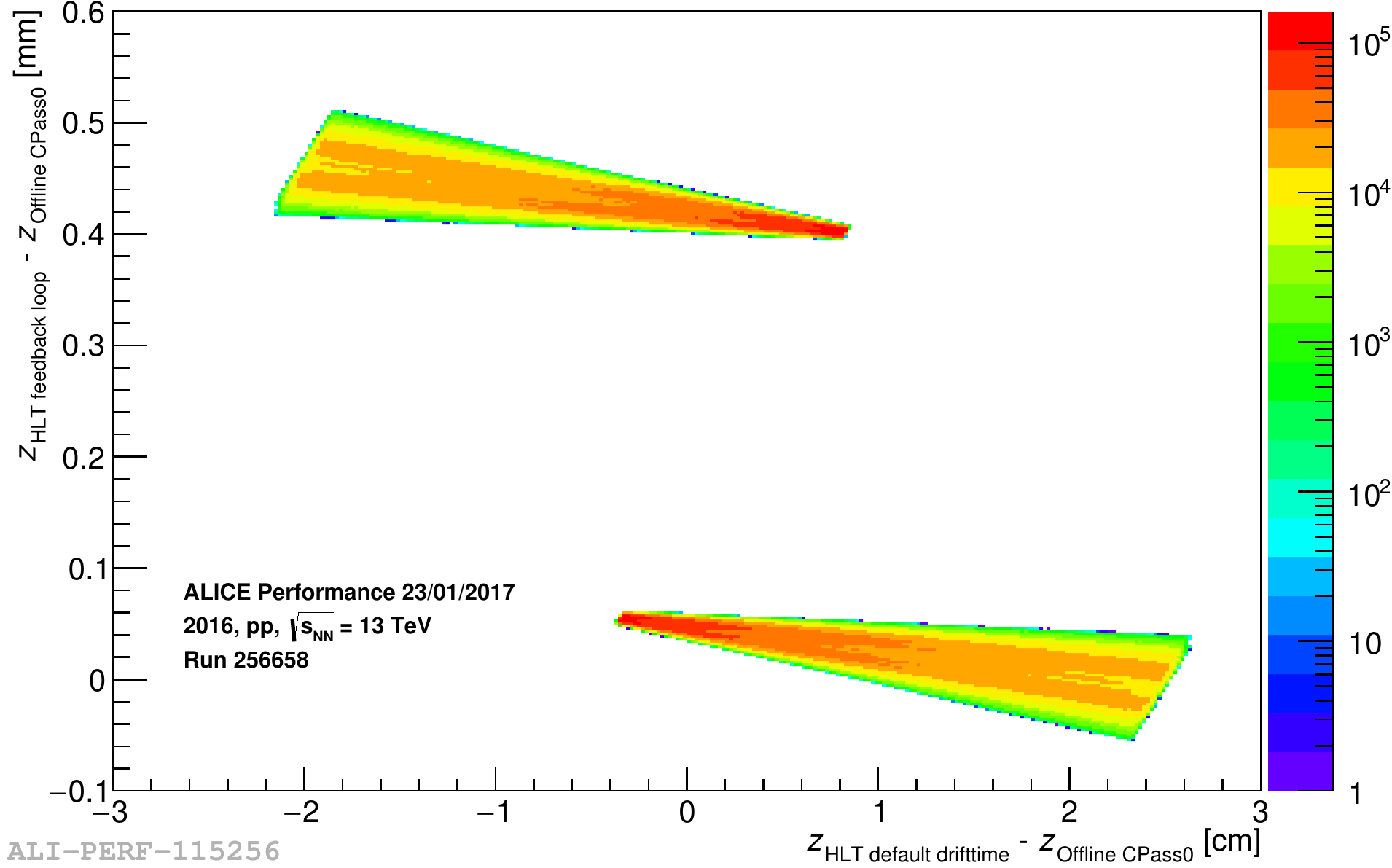}
\caption{Comparison of the cluster~$z$-coordinate deviation using online calibration ($y$-axis) to an uncalibrated HLT ($x$-axis).
Similar to Figure~\ref{fig_calib_dz}, but here also the~$x$-axis shows a difference of cluster positions, comparing the new online calibration and no calibration.}
\label{fig_calib_feedback_loop}
\end{figure}

Figure~\ref{fig_calib_feedback_loop} compares the improvements in the HLT cluster position with online calibration to the situation before when HLT had to rely only on a default drift time.
The figure shows again on a per-cluster level the deviation from the HLT cluster coordinates to the offline cluster coordinates after CPass 0.
The x-axis shows the deviation without online calibration, where the cluster positions are off by up to~$2.5$\,cm.
Overall accuracy of cluster positions in the HLT thus improves by the factor~50.

\begin{figure}[!t]
\centering
\includegraphics[width=3.4in]{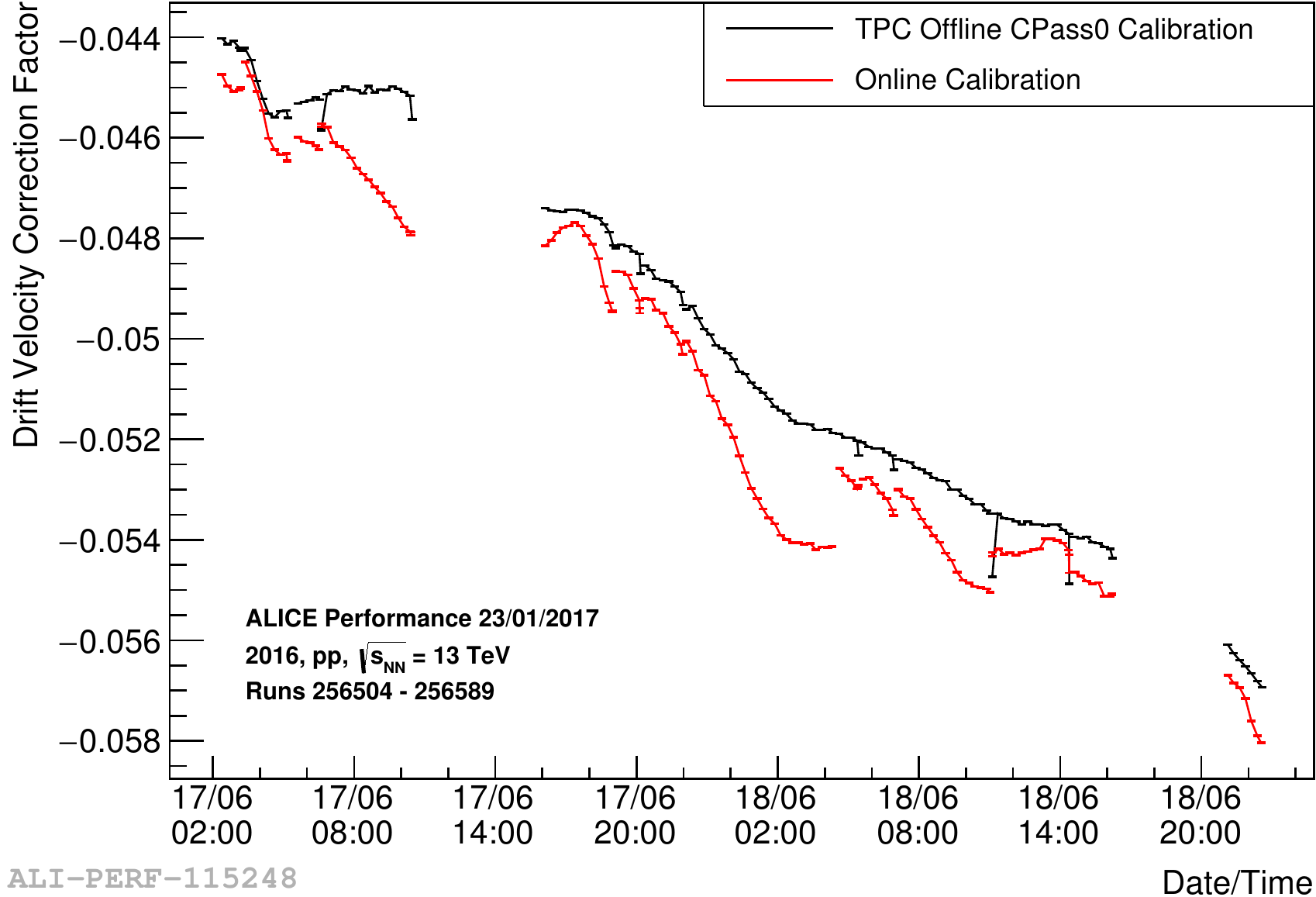}
\caption{Drift velocity correction factor ($1 + \alpha$) computed by offline CPass0 (black, upper curve) and by online calibration (red, lower curve).}
\label{fig_drift_time}
\end{figure}

Finally, we cross-check the results by comparing the drift velocity correction factors~$\alpha$ from CPass 0 and from online calibration directly in Figure~\ref{fig_drift_time}.
The figure shows the HLT and CPass 0 correction factors for several data-taking runs on two consecutive days.
We mention that the curves are not expected to overlap completely because some configuration objects (e.\,g.~pressure and temperature) differ.
CPass 0 always has the correct information.
HLT used a temperature object from few days earlier than the first run in the figure.
We have checked using data replay in the HLT with the correct temperature object that the incorrect temperature causes exactly the offset.
The pressure parameter for the HLT is only updated at the beginning of a run.
Since the environment parameters in the HLT are constant during a run, the correction factor accommodates for changes in pressure and temperature.
As the offline CPass 0 uses the correct time development, the two lines for the correction factors diverge during a run.
The HLT curve is then reset at the beginning of the next run when parameters are updated.
The gaps in the figures are periods when LHC delivered no stable beam.
Overall, the figure shows that the online drift time correction factors follows the trend of the offline factor.
It has an offset and it shows some artifacts which are caused by different temperature and pressure information.

\section{Dependencies of Calibration Objects and Additional Calibration Tasks}

\label{sec:dep}

Offline analysis runs multiple reconstruction passes over the data.
The calibration objects within a pass remain the same.
In the HLT, there is only one real-time pass over the data while the calibration objects change: they are updated by the feedback loop.
This introduces an additional level of complexity.

For instance, the fits in the TPC drift time calibration are based on the misalignment of TPC and ITS tracks.
After the feedback loop has reinserted the first calibration objects into the reconstruction, the TPC clusters are calibrated and TPC and ITS tracks should be aligned.
In theory, this could be treated in the calibration procedure, if the TPC drift time calibration would take into account the actual non-constant calibration object used in the reconstruction.
This is not supported at the moment.
Hence, the HLT must only use clusters in the calibration task that are transformed with the default calibration object.
In contrast, we want to use calibrated clusters for the tracking.
The TPC drift time calibration does not use the fitted parameters of the track itself, but it does a refit of the track.
Therefore, the HLT currently creates two instances of the clusters.
One instance uses the feedback loop to create calibrated clusters for track reconstruction.
The second instance always uses the default calibration and is currently used in the calibration.
In other words, we produce both the clusters as in CPass 0 and as in the next pass at the same time.
These two instances are seen in Figure~\ref{fig_hlt_overview}.
Since we ensure consistent cluster indices, the calibration can use the default clusters to refit the tracks in the way it needs.
Due to this refit, it is not necessary to run the track finding twice.

These two instances of the clusters match more or less the different offline passes: CPass 0 for the calibration and an additional pass for the real reconstruction.
However, offline has two subsequent calibration passes before the physics pass.
In order to run CPass 1 calibration tasks in the HLT, we will have to mimic the 3 pass behavior of offline in the HLT.
This probably means that we will need another instance of the data.
The data reconstructed with calibrated clusters needs to be fed into a second calibration component that runs the CPass 1 calibration tasks.
The CPass 1 calibration objects can be fed back in the same way as the TPC drift time calibration.
Based on that the HLT could run the fully calibrated reconstruction.
In order to save compute resources, we will have to make sure not to execute too many reconstruction steps multiple times in parallel for the same event.

\section{Summary and Conclusions}

We have presented a scheme for online calibration in the HLT.
By wrapping the offline analysis task inside an HLT component, we can in principle run all offline tasks.
The only prerequisite is that the task has to use HLT's new flat event data structure format, which speeds up the transfer of data structures between HLT components.
We have exemplarily adapted the TPC drift time calibration task accordingly.

The produced calibration objects can be made available to HLT reconstruction via a feedback loop.
This improves the HLT reconstruction and QA capabilities greatly.
The online calibration with feedback loop approach in the HLT is slightly different from the offline calibration approach.
Since the HLT cannot cache all the data and postpone the reconstruction until the calibration has finished, the feedback loops operates with a pipeline.
Calibration objects are produced, updated, and made available to the reconstruction on a regular basis.
During the very beginning of the run, the reconstruction still has to rely on the default calibration when no updated calibration is available yet.
As soon as the first calibration object is passed back to the reconstruction via the feedback loop, the HLT operates with calibration.
This imposes a time constraint to the calibration task: the update interval must be shorter than the stability interval of the calibration.
The initial phase without calibration is usually short compared to the duration of the entire run.
The produced calibration objects can be used for offline reconstruction, and in that case they cover the entire run.

Several new HLT features facilitate the online calibration.
Besides the new flat data structure for faster data exchange, the HLT provides now asynchronous processing capabilities for long-running task.
These are decoupled from the normal HLT operation in individual processes such that even fatal errors do not affect data taking.
We use a new ZeroMQ transport model for the feedback loop.
A new fast standalone HLT ITS reconstruction decouples ITS from TPC tracking and ensures that there is no bias in the TPC-ITS matching.
Using two instances of the cluster transformation, we can run the tracking with calibrated clusters and the online calibration task in parallel.

We did a first successful test of online calibration with feedback loop during lead-lead data taking in December 2015.
This is the highest conceivable load scenario when ALICE was taking lead-lead data at design luminosity.
The HLT managed to operate online TPC drift time calibration and feedback loop with a maximum delay of~$9.2$ minutes which lies in the stability interval of~$15$ minutes.
During that test, the HLT could run the calibration task for 31 events per second, and by dedicating more compute resources we could improve the rate to~$81$\,Hz.
In the meantime, we have improved the implementation, and the HLT is currently running the online calibration in 2016 proton-proton data taking requiring less compute resources.
During the lead-lead run in 2015, the 600 CPU cores running the online calibration matched~277.5 tracks per second in average.
The improved software needs 510 CPU cores for matching 480 tracks per second in the current 2016 proton-proton run.

One deficiency at the moment is the lack of updated, recent configuration information from temperature and pressure sensors, but also other relevant conditions like the TPC high voltage.
We demonstrated that, in principle, the calibration process compensates for this, resulting in only slightly worse cluster~{$z$-coordinates} differing by less than~$1.5$\,mm in the worst case.
The more severe caveat is the fact that the produced calibration object is linked to the configuration objects used during the calibration, i.\,e.~the calibration objects produced by the HLT need to be operated using the same outdated temperature information that were used in the HLT during the calibration task.
Therefore, the HLT objects cannot be used in offline reconstruction directly yet.
For the future we want to eliminate this limitation.
We will have to make the TPC temperature and pressure information available to the HLT in real time.
In addition, the HLT needs to update the configuration objects with new sensor information periodically to have time-dependent sensor graphs.
This update can happen through the same mechanism that updates the drift time object containing the cluster transformation described in Section~\ref{sec_loop}.

The current drift time objects produced by the HLT show the same trends as the objects produced by offline CPass 0.
The actual drift time is different by a certain factor stemming from different temperature and pressure configurations used.
With the online calibration, the difference of cluster~{$z$-position} in HLT and offline is less than~$0.5$\,mm.
This demonstrates the quality of HLT TPC and ITS tracking and is a great improvement over the previous HLT reconstruction relying on default drift time without online calibration.
Before, the HLT clusters were off by up to~$2.5$\,cm in~{$z$-direction}.

Having a general wrapper that can run any analysis task based on the flat data structure allows us to adopt other calibration tasks with little effort.
We have already started to adapt the gain calibration for the dE/dx calibration accordingly.
Naturally, the concept is not limited to calibration tasks.
For instance, we are working on an adaptation of the TPC QA task for the HLT.
In the long run, all these new developments are mandatory for the online-offline compute concept~O$^2$~\cite{bib:o2} foreseen for ALICE Run 3 after the next long shutdown.
In this future scenario data compression will rely on precise reconstruction which makes online calibration a necessity.
Through the current developments and improvements we already gather necessary experience for this new scenario.


\end{document}